\documentstyle[a4wide]{article}
\begin{document}

\newcommand{\lsim}{\mbox{\raisebox{-.9ex}{~$\stackrel{\mbox{$<$}}{\sim}$~}}}
\newcommand{\gsim}{\mbox{\raisebox{-.9ex}{~$\stackrel{\mbox{$>$}}{\sim}$~}}}

\begin{titlepage}

\title{\Large\bf Inflation at the TeV 
scale with 
a PNGB curvaton}

\author{Konstantinos Dimopoulos\\
\mbox{\hspace{1cm}}\\
{\it Physics Department, Lancaster University, Lancaster LA1 4YB, U.K.}\\
\mbox{\hspace{1cm}}}

\date{October, 2005}

\maketitle

\abstract{ \normalsize\noindent 
We investigate a particular type of curvaton mechanism, under which
inflation can occur at Hubble scale of order 1~TeV. The curvaton is a 
pseudo Nambu-Goldstone boson, whose order parameter increases after a phase
transition during inflation, triggered by the gradual decrease of the Hubble 
scale. The mechanism is studied in the context of modular inflation, where
the inflaton is a string axion. We show that the mechanism is successful
for natural values of the model parameters, provided the phase transition
occurs much earlier than the time when the cosmological scales exit the 
horizon. Also, it turns our that the radial mode for our curvaton 
must be a flaton field.}

\thispagestyle{empty}

\end{titlepage}

\pagebreak

%
%
Inflation is the only compelling theory to date for the solution of the
horizon and flatness problems of the big bang cosmology as well as for
explaining structure formation in the Universe. Recent precise observations 
have confirmed the basic predictions of the inflationary paradigm by 
ascertaining the spatial flatness of the Universe and the approximate scale 
invariance of the density perturbations, which give rise to the 
anisotropy of the Cosmic Microwave Background Radiation (CMBR) and seed 
structure formation. These exciting developments have rendered the 
inflationary paradigm a necessary extension to the hot big bang standard 
cosmology.

In the light of precision data, inflation model-building can be upgraded beyond
the simple single-field stage of its early beginnings. Indeed, more complex and
realistic models of inflation, with tighter connections to the theory, less 
fine tunning and enhanced predictability and falsifiability are now possible to
construct, making use of the rich content of particle physics. A first such 
example is the well-known hybrid inflation model \cite{hybrid}, which couples
the inflaton field to the Higgs field of a Grand Unified Theory (GUT) in
order to obtain without tunning the desired false vacuum energy scale
\cite{susyhybrid}. In hybrid inflation the inflationary period is terminated
through the dynamics of this other field. 

In an analogous manner, one can attribute the generation of density 
perturbations during inflation to a field other than the inflaton \cite{LW}. 
This so-called curvaton field allows inflation to take place at a much 
lower energy scale than the typically required GUT-scale \cite{liber} and,
in general, may relax a number of constraints regarding inflation
model-building \cite{liber2}. Low-scale inflation can revamp a number of
inflation models that are well motivated on particle physics grounds
\cite{liber}. It is important to stress here that the curvaton is {\em not}
an ad~hoc additional degree of freedom introduced ``by hand'', but it may
be a realistic field, already present in simple extensions of the standard 
model. Indeed, many such examples exist in the literature 
\cite{curv,pngb}.

However, even when a curvaton field is considered, there exists a lower bound
for the inflationary scale, which, for generic curvaton models, can be quite
tight \cite{curvbound}. This lower bound can be substantially relaxed for 
certain types of curvaton models \cite{amplif}, which enables inflation to be 
directly connected to realistic, beyond the standard model physics. 

In this letter I present a curvaton model which allows inflation at a Hubble 
scale as low as 1~TeV. The curvaton field is a pseudo Nambu-Goldstone boson
(PNGB), whose order parameter is substantially increased after the cosmological
scales exit the horizon during inflation. As shown in \cite{amplif}, the result
of this increase is to amplify the curvaton's perturbations. This enables even
low-scale inflation to generate density perturbations of the observed 
amplitude. In the curvaton model presented, the increase of the PNGB order
parameter follows a phase transition during inflation, which releases the
radial mode from the top of the potential hill. 

The use of a PNGB curvaton is highly motivated because such a curvaton can be
naturally light during inflation, since its mass is protected by the global
U(1) symmetry \cite{pngb}. This dispenses with the danger imposed by
supergravity corrections, which typically lift the flatness of the scalar
potential \cite{randall}.
We investigate the performance of the curvaton model in the context of modular 
inflation, which corresponds to Hubble scale of order 1~TeV. Modular inflation
is a well motivated model, which uses a string axion as the inflaton 
\cite{modular}. 

%
%

Let us begin by presenting the amplification mechanism for the curvature 
perturbations.
We discuss here the case of an PNGB curvaton, whose order parameter has a 
different (larger) expectation value in the vacuum than during inflation and, 
in particular, when the cosmological scales exit the horizon. Thus, the 
potential for the curvaton field $\sigma$ is
\begin{eqnarray}
V(\sigma)=(v\tilde{m}_\sigma)^2[1-\cos(\sigma/v)] & \Rightarrow & 
V(|\sigma|<v)\simeq\frac{1}{2}\tilde{m}_\sigma^2\sigma^2,
\label{Vs}
\end{eqnarray}
where \mbox{$v=v(t)$} is the order parameter determined by the expectation 
value of the radial field $|\phi|$  and
\mbox{$\tilde{m}_\sigma=\tilde{m}_\sigma(v)$} is the mass of the curvaton at a 
given moment. In the true vacuum we have \mbox{$v=v_0$} and 
\mbox{$\tilde{m}_\sigma=m_\sigma$} with $v_0$ being the vacuum expectation 
value (VEV) of the radial 
field and $m_\sigma$ being the mass of the curvaton in the vacuum.


Let us
demonstrate that the curvaton perturbations can be
amplified by the non-trivial evolution of the radial field.
We begin by using the fact that \cite{LW}:
\begin{equation}
\zeta\sim\Omega_{\rm dec}\zeta_\sigma\;,
\label{zeta}
\end{equation}
where \mbox{$\zeta\simeq\sqrt{{\cal P}_\zeta}=2\times 10^{-5}$} is the 
curvature perturbation of the 
Universe, \mbox{$\Omega_{\rm dec}\leq 1$} is the density fraction of the 
curvaton density over the density of the Universe at the time of the decay of 
the curvaton:
\begin{equation}
\Omega_{\rm dec}\equiv\left.\frac{\rho_\sigma}{\rho}\right|_{\rm dec}\leq 1
\label{r}
\end{equation}
and $\zeta_\sigma$ is the curvature perturbation of the curvaton field
$\sigma$, which is given by
\begin{equation}
\zeta_\sigma\sim\left.\frac{\delta\sigma}{\sigma}\right|_{\rm dec}\sim
\left.\frac{\delta\sigma}{\sigma}\right|_{\rm osc}\;,
\label{zs1}
\end{equation}
where `osc' denotes the time when the curvaton oscillations begin. 
Note that, non-gaussianity constraints from the observations from the WMAP 
satellite \cite{nongauss} restrict the range of 
$\Omega_{\rm dec}$ as follows:
\begin{equation}
10^{-2}\leq\Omega_{\rm dec}\leq 1\,.
\label{wmap}
\end{equation}

In this paper we consider the inflationary Hubble scale to be comparable to 
the tachyonic mass of the radial field, which determines the value of the order
parameter of our PNGB curvaton. This means that the evolution of the radial 
field ceases at (or soon after) the end of inflation. Therefore, at the end of
inflation, \mbox{$v\rightarrow v_0$} and the mass of the curvaton assumes
its vacuum value $m_\sigma$. Hence, in the following {\em we assume that 
the curvaton mass has already assumed its vacuum value before the onset of 
the curvaton oscillations.} Consequently, the curvaton oscillations begin when
\begin{equation}
H_{\rm osc}\sim m_\sigma\;.
\label{Hosc}
\end{equation}
Before the oscillations begin the curvaton
is overdamped and remains frozen. This means that
\mbox{$\theta_{\rm osc}\simeq\theta_*$},
where the `*' denotes the values of quantities at the time when the 
cosmological scales exit the horizon during inflation and 
\begin{equation}
\theta\equiv\sigma/v\;,
\label{theta}
\end{equation}
with \mbox{$\theta\in(-\pi,\pi]$}. Hence, for the curvaton fractional 
perturbation we find
\begin{equation}
\left.\frac{\delta\sigma}{\sigma}\right|_*
=\left.\frac{\delta\theta}{\theta}\,\right|_*\simeq
\left.\frac{\delta\sigma}{\sigma}\right|_{\rm osc}.
\label{fractional}
\end{equation}
Now, for the perturbation of the 
curvaton we have
\begin{equation}
\delta\sigma_*=\frac{H_*}{2\pi}\,,
\label{dsH}
\end{equation}
We assume that the expectation value of the radial field during inflation is 
smaller compared to its VEV by a factor
\begin{equation}
\varepsilon\equiv\frac{v_*}{v_0}\ll 1\,.
\label{eps}
\end{equation} 
Combining Eqs.~(\ref{fractional}), (\ref{dsH}) and (\ref{eps}), in view also of
Eq.~(\ref{theta}), we find
\begin{equation}
\delta\sigma_{\rm osc}\simeq\frac{H_*}{2\pi\varepsilon}\,,
\label{dsosc}
\end{equation}
which means that after the end of inflation, when the radial field assumes its 
VEV, {\em the curvaton perturbation is amplified by a factor} 
$\varepsilon^{-1}$ (see Figure~1). From Eqs.~(\ref{zeta}) and (\ref{zs1}) 
we have
\begin{equation}
\sigma_{\rm osc}\sim (\Omega_{\rm dec}/\zeta)\,\delta\sigma_{\rm osc}\;.
\end{equation}
Using Eqs.~(\ref{dsH}) and (\ref{dsosc}), we can recast the above as
\begin{equation}
\sigma_{\rm osc}\sim\frac{H_*\Omega_{\rm dec}}{\pi\varepsilon\zeta}\,.
\label{sosc}
\end{equation}
We may obtain a lower bound on $\varepsilon$ as follows:
\begin{equation}
\frac{\delta\sigma_*}{\sigma_*}\leq 1\quad\Rightarrow\quad
\varepsilon\geq\varepsilon_{\rm min}\equiv\frac{H_*}{2\pi v_0}\,,
\label{epsbound}
\end{equation}
where we have used Eqs.~(\ref{theta}), (\ref{dsH}) and (\ref{eps}) and that 
\mbox{$\sigma_{\rm osc}\lsim v_0$}.

\input{epsf}

\begin{center}
\begin{figure}

\begin{picture}(10,10)
\put(60,-200){

\leavevmode
\hbox{%
\epsfxsize=4in
\epsffile{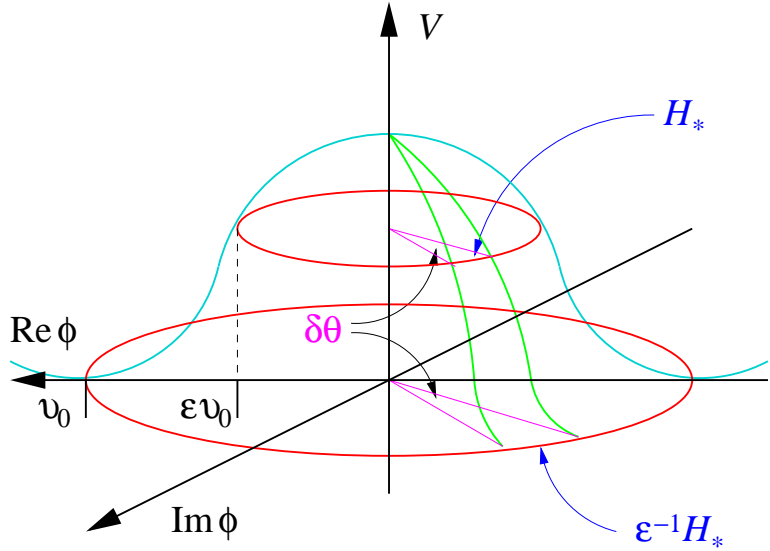}}
}



\end{picture}

\vspace{7.5cm}

\caption{
Schematic representation of the amplification of the PNGB curvaton 
perturbation, when the order parameter $v$ increases from the value it has
when the cosmological scales exit the horizon \mbox{$v_*=\varepsilon v_0$} to 
its vacuum value $v_0$. The perturbation at horizon crossing has amplitude
\mbox{$\delta\sigma_*\sim H_*$}, which corresponds to a phase perturbation for
the radial field $|\phi|$ of magnitude 
\mbox{$\delta\theta=\delta\sigma_*/v_*$}.
As the order parameter grows $\delta\theta$ remains constant (the phase
perturbation is frozen on superhorizon scales) but the amplitude of the 
curvaton perturbation is increased up to 
\mbox{$\delta\sigma\sim\varepsilon^{-1}H_*$}.
}
\end{figure}
\end{center}


Now, as is shown in \cite{amplif}, in the case when the curvaton oscillations
begin after the radial field has attained its VEV, we have\footnote{%
We use natural units, where \mbox{$c=\hbar=1$} and 
Newton's gravitational constant is \mbox{$G=8\pi m_P^{-2}$} with 
\mbox{$m_P=2.44\times 10^{18}$GeV} being the reduced Planck mass.}
\begin{equation}
H_*\sim\Omega_{\rm dec}^{-2/5}
\left(\frac{H_*}{\min\{m_\sigma, \Gamma_{\rm inf}\}}\right)^{1/5}
\left(\frac{\max\{H_{\rm dom}, \Gamma_\sigma\}}{H_{\sc bbn}}\right)^{1/5}
(\pi\varepsilon\zeta)^{4/5}
(T_{\sc bbn}^2m_P^3)^{1/5},
\label{H*}
\end{equation}
where $\Gamma_{\rm inf}$ and $\Gamma_\sigma$ are the decay rates of 
the inflaton and the curvaton fields respectively, $H_{\rm dom}$ is the Hubble
parameter at the time when the curvaton density dominates the Universe (if the
curvaton does not decay earlier) and
\mbox{$H_{\sc bbn}\sim T_{\sc bbn}^2/m_P$} is
the Hubble parameter at the time of Big Bang Nucleosynthesis (BBN), with 
\mbox{$T_{\sc bbn}\sim 1$ MeV}.

Now, we require that the curvaton 
decays before BBN, i.e. \mbox{$\Gamma_\sigma>H_{\sc bbn}$}. We also have
\mbox{$\Gamma_{\rm inf}\leq H_*$}. Hence, Eq.~(\ref{H*}) 
provides the following bound
\begin{equation}
H_*>\Omega_{\rm dec}^{-2/5}
(\pi\varepsilon\zeta)^{4/5}(T_{\sc bbn}^2m_P^3)^{1/5}\sim
(\varepsilon^2/\Omega_{\rm dec})^{2/5}
\times 10^7{\rm GeV}\,.
\label{H*bound}
\end{equation}

Furthermore, we also note that
\begin{equation}
\Gamma_\sigma\geq\frac{m_\sigma^3}{m_P^2}\,,
\label{Ggrav}
\end{equation}
where the equality corresponds to gravitational decay.
The above can be shown \cite{amplif} to imply that
\begin{equation}
H_*\geq\Omega_{\rm dec}^{-1}(\pi\varepsilon\zeta)^2m_P
\left(\frac{m_\sigma}{H_*}\right)
\max\left\{1, \frac{m_\sigma}{\Gamma_{\rm inf}}\right\}^{1/2},
\label{Hbound1}
\end{equation}
which results in the bound
\begin{equation}
H_*\geq\Omega_{\rm dec}^{-1}
(\pi\varepsilon\zeta)^2m_P\left(\frac{m_\sigma}{H_*}\right)\sim
(\varepsilon^2/\Omega_{\rm dec})
\times 10^{10}{\rm GeV}\left(\frac{m_\sigma}{H_*}\right).
\label{Hbound}
\end{equation}
This bound
may be relaxed if $\varepsilon$ is small enough. In particular,
for a PNGB curvaton we may have \mbox{$m_\sigma <H_*$}. 
Comparing the bound in Eq. (\ref{H*bound}) 
with the one in Eq. (\ref{Hbound}) 
we find that the former
bound is more stringent if
\begin{equation}
\varepsilon<\frac{1}{\pi\zeta\sqrt{\Omega_{\rm dec}}}
\left(\frac{T_{\sc bbn}}{m_P}\right)^{1/3}
\left(\frac{H_*}{m_\sigma}\right)^{5/6}\sim
10^{-3}\,\Omega_{\rm dec}^{-1/2}
(H_*/m_\sigma)^{5/6}
\end{equation}
Thus, for \mbox{$\varepsilon\ll 1$}, the second
bound is typically less stringent than the first one. 



During inflation, the evolution of 
the order parameter of the PNGB curvaton, is subject to an 
important constraint, which has to do with preserving the scale invariance of 
the spectrum of the curvature perturbations.

The amplitude of the density perturbations is determined by the magnitude of 
the perturbations of the curvaton field, which, in this scenario, apart 
from the scale of $H_*$ is also determined by the amplification factor 
$\varepsilon^{-1}$. The latter is determined by the value of the order 
parameter $v_*$ when the curvaton quantum fluctuations exit the horizon during 
inflation. A strong variation of $v(t)$ at that time results in a strong 
dependence of $\varepsilon(k)$ on the comoving momentum scale $k$, which would 
reflect itself on the perturbation spectrum threatening significant 
departure from scale invariance. 

In Ref.~\cite{amplif} is was shown that, in order for this to be avoided, 
the rate of change of the radial field must be constrained as
\begin{equation}
|\dot v/v|_*=|\dot\phi/\phi|_*\ll H_*\;,
\label{nsbound}
\end{equation}
where $|\phi|$ is the radial field, which determines the value of the order 
parameter. In fact, the contribution to the spectral index due to
the evolution of $v$ is \mbox{$\delta n_s=-2H_*^{-1}(\dot v/v)_*$}.
From the above it is evident that, in order not to violate the 
observational constraints regarding the scale invariance of the density 
perturbation  spectrum, {\em the roll of the radial field has to be at most 
very slow when the cosmological scales exit the horizon}. However, this cannot 
remain so indefinitely because we need \mbox{$v_0\gg v_*$} 
to have substantial amplification of the perturbations (i.e. 
\mbox{$\varepsilon\ll 1$}).
Consequently, $v$ has to increase dramatically 
at some point {\em after} the exit of the cosmological scales from the horizon.
This requirement is crucial for model-building.\footnote{The requirement in
Eq.~(\ref{nsbound}) may be even more fundamental in origin. Indeed, a PNGB
with rapidly varying order parameter cannot be treated as an effectively
free field. I would like to thank D.H.~Lyth for pointing this out.}

In our model, we will show that the evolution of $v$ begins at a phase 
transition during inflation. Initially, the growth of $v$ is
very slow, but later, near the end of inflation, $v$ grows substantially until 
it reaches its vacuum value $v_0$.



%
%

Let us now briefly describe the model of inflation.
We are going to consider a PNGB curvaton $\sigma$ whose radial field $|\phi|$ 
is of bare mass similar to the Hubble parameter during inflation, that is
\begin{equation}
m_\phi\sim H_*\;.
\label{mphi}
\end{equation}
This has the advantage that the radial field rolls substantially by the end of
inflation so that $\varepsilon$ can be very small. In particular, we will 
assume that the tachyonic mass of the radial field is a soft mass generated by 
supersymmetry breaking and it is, therefore, roughly of the electroweak scale
$m_{3/2}$. Hence, we consider inflation at the intermediate scale
\begin{equation}
V_*^{1/4}\sim\sqrt{m_{3/2}m_P}\sim 10^{10.5}{\rm GeV}
\quad\Rightarrow\quad H_*\sim m_{3/2}\;.
\end{equation}

A particular example of such an inflation model 
(but, by all means, not the only one) is modular inflation \cite{modular}, 
where the inflaton field $s$ is a string axion, whose flatness 
is lifted by gravity mediated supersymmetry breaking. 
In this model the inflationary potential is of the form:
\begin{equation}
V(s)=V_{\rm inf}-\frac{1}{2}m_s^2s^2+\cdots\;,
\label{Vinf}
\end{equation}
where the ellipsis denotes terms, which are expected to stabilise the potential
at \mbox{$s_{\sc vev}\sim m_P$}. Therefore, in the above we have
\begin{equation}
V_{\rm inf}\sim (m_{3/2}m_P)^2 \qquad {\rm and} \qquad 
m_s\sim H_{\rm inf}\sim m_{3/2}\;,
\label{H0}
\label{V0}
\end{equation}
where \mbox{$H_{\rm inf}\simeq\sqrt{V_{\rm inf}}/3m_P$}.

This inflation model results in fast roll inflation \cite{fastroll}, where 
\begin{equation}
s=s_{\rm in}\exp(F_s\Delta N) \qquad {\rm and} \qquad 
F_s\equiv\frac{3}{2}\left(\sqrt{1+4c/9}-1\right)
 \qquad {\rm with} \qquad c\equiv\left(\frac{m_s}{H_{\rm inf}}\right)^2
\sim 1\,,
\label{Fs}
\end{equation}
where $\Delta N$ is the number of the elapsed e-foldings.
From the above one can easily obtain the inflation scale 
$N$ e-foldings before the end of inflation, as
\begin{equation}
V(N)\simeq V_{\rm inf}\left(1-e^{-2F_sN}\right)\,.
\label{VN}
\end{equation}

Even though fast-roll, modular inflation keeps
the Hubble parameter $H$ rather rigid. Indeed, it can be easily shown that
\begin{equation}
\epsilon=
\frac{1}{2}F_s^2\left(\frac{s}{m_P}\right)^2\simeq\frac{1}{2}F_s^2 e^{-2F_sN}
\ll 1\,,
\label{vareps}
\end{equation}
because \mbox{$F_s\sim 1$} and \mbox{$s\ll m_P$} during inflation, with 
\mbox{$\epsilon\ll 1$} being one of the so-called slow roll parameters
defined as
\begin{equation}
\epsilon\equiv-\frac{\dot H}{H^2}\,.
\label{SR}
\end{equation}

For modular inflation the initial conditions for the inflaton field
are determined by the quantum fluctuations, which send the field off the top
of the potential hill. (The modulus can be considered to be originally placed
at the local maximum because the latter can be thought of as a fixed point of 
the symmetries.) Hence, we expect that the initial value for the inflaton is
\begin{equation}
s_{\rm in}\simeq
H_{\rm inf}/2\pi\,.
\label{s0}
\end{equation}
Using the above and considering that the final value is 
\mbox{$s_{\sc vev}\sim m_P$}, we can estimate, through the use of 
Eq.~(\ref{Fs}), the total number of e-foldings as
\begin{equation}
N_{\rm tot}\simeq\frac{1}{F_s}
\ln\left(\frac{m_P}{m_{3/2}}\right),
\label{Ntot}
\end{equation}
where we took into account Eq.~(\ref{H0}).

Let us turn our attention to the curvaton model.
%
%
Consider 
the superpotential\footnote{Such type of superpotential is reminiscent of
supersymmetric realisations of the Peccei-Quinn symmetry in which the 
Peccei-Quinn scale is generated dynamically \cite{PQ}.}
%
\begin{equation}
W=\frac{\lambda}{n+3}\frac{\phi^{n+3}}{m_P^n}\;,
\label{W}
\end{equation}
where \mbox{$n\geq 0$} and
the complex field $\phi$ can be thought to contain the curvaton phase 
field $\sigma$ and one radial field $|\phi|$ as follows:
\begin{equation}
\phi\equiv|\phi|e^{i\theta}=|\phi|\exp(\sigma/\sqrt 2 v)\,.
\label{absphi}
\end{equation}
Then the scalar potential can be written as
\begin{eqnarray}
 V & = & (C_\phi H^2-m_\phi^2)|\phi|^2+\left[(C_A H+A)\frac{\lambda}{n+3}
 \frac{\phi^{n+3}}{m_P^n}+h.c.\right]+\lambda^2\frac{|\phi|^{2n+4}}{m_P^{2n}}
=\nonumber\\
 & = & (C_\phi H^2-m_\phi^2)|\phi|^2+\lambda^2\frac{|\phi|^{2n+4}}{m_P^{2n}}+
(C_A H+A)\frac{2\lambda}{n+3}\frac{|\phi|^{n+3}}{m_P^n}\cos[(n+3)\theta]
\label{V}
\end{eqnarray}
where $m_\phi$ and $A$ are soft supersymmetry breaking mass-scales at
zero temperature, both given by the electroweak scale $m_{3/2}$. 
Note that we have put negative mass-squared for the
$|\phi|$ field at zero temperature to break the U(1) symmetry.
We also considered corrections coming from supergravity, which provide 
effective mass terms of order $H$ \cite{randall} 
(for their effect on curvaton physics see Ref.~\cite{CD}). 
Absorbing the $(n+3)$ factor into $\theta$ (and shifting the latter 
by $\pi$) \footnote{This, in effect, means considering the range:
\mbox{$-\frac{\pi}{n+3}<\theta-\pi\leq\frac{\pi}{n+3}$}.}
we can write the curvaton potential as:
\begin{equation}
V(\sigma) \approx \lambda(C_A H + A ) v^3\left(\frac{v}{m_P}\right)^n\,
\left[1-\cos\left(\frac{\sigma}{v}\right)\right]\,.
\label{Vcurv}
\end{equation}

We are going to assume that the U(1)
symmetry is broken at some moment during inflation with
\mbox{$H_*\sim m_\phi\sim m_{3/2}$}. Hence we take \mbox{$C_\phi\sim + 1$}. 
After this moment the radial field $|\phi|$ begins to grow, which can result 
in \mbox{$\varepsilon\ll 1$}. In time, after the symmetry breaking,
the tachyonic effective mass of the radial field approaches its vacuum value
$m_\phi$ as the supergravity correction diminishes due to the gradual decrease 
of the Hubble parameter. 

After the phase transition, the time-dependent minimum of the potential
of the radial field is given by
\begin{eqnarray}
|\phi|_{\rm min} & = & 
\left(\lambda^{-1}m_P^n\sqrt{m_\phi^2-C_\phi H^2}\right)^{\frac{1}{n+1}},
\label{vexact}
\end{eqnarray}
which gradually grows. Soon $|\phi|_{\rm min}$ assumes its vacuum value:
\begin{equation}
v_0\sim 
\left(\lambda^{-1}m_P^nm_\phi\right)^{\frac{1}{n+1}},
\label{fPQ}
\end{equation}
From the above and also in view of Eqs.~(\ref{Vs}) and (\ref{Vcurv}) we find
%
\begin{equation}
\tilde{m}_\sigma^2\approx\lambda(C_A H+A)v\left(\frac{v}{m_P}\right)^n.
\label{mcurv}
\end{equation}
Evaluating the above after the order parameter assumes its vacuum value
\mbox{$v\rightarrow v_0$} we obtain
\begin{equation}
m_\sigma^2\approx(C_AH+A)m_\phi\,,
\end{equation}
where we used Eq.~(\ref{fPQ}).
Since \mbox{$m_\phi\sim A\sim m_{3/2}$}, \mbox{$C_A\sim 1$} and 
\mbox{$H\leq H_*$} we find that
\begin{equation}
m_\sigma\sim m_{3/2}\sim H_*\;.
\end{equation}

However, during inflation the effective mass of the curvaton is much smaller.
Indeed, in view of Eq.~(\ref{mcurv}), 
we get
\begin{equation}
\frac{\tilde m_\sigma^2}{m_\sigma^2}\sim\left(\frac{v}{v_0}\right)^{n+1}
\quad\Rightarrow\quad
\tilde m_\sigma(v_*)
\sim\varepsilon^{\frac{n+1}{2}}m_\sigma\;,
\label{mH}
\end{equation}
where we used Eq.~(\ref{eps}). Therefore, since \mbox{$\varepsilon\ll 1$} and 
\mbox{$m_\sigma\sim m_{3/2}\sim H_*$} we see that, during inflation
\mbox{$\tilde m_\sigma\ll H_*$}, i.e. the PNGB is appropriately light and
can act as a curvaton field.



%
%

Let us now calculate the value of $\varepsilon$ required so that the scenario 
works. Firstly, we note that, in our case, the curvaton assumes a random value
at the phase transition, which typically is 
\mbox{$\sigma\sim v$}.
After the end of inflation and 
before the onset of the oscillations the field is overdamped and remains 
frozen. Hence, we expect that at the onset of the oscillations we have:
\begin{equation}
\sigma_{\rm osc}\sim \theta\, v_0\;,
\label{soscv0}
\end{equation}
where, typically, \mbox{$\theta\sim{\cal O}(1)$} and
we took into account that the radial field assumes its VEV very soon 
after the end of inflation. 
Combining Eqs.~(\ref{sosc}) and (\ref{soscv0}),
we find
\begin{equation}
\varepsilon\sim\frac{\Omega_{\rm dec}}{\pi\zeta\theta}
\left(\frac{m_{3/2}}{m_P}\right)^{\frac{n}{n+1}},
\label{epszeta}
\end{equation}
where we also used Eq.~(\ref{fPQ}) taking \mbox{$m_\phi\sim m_{3/2}$} and 
\mbox{$\lambda\sim 1$}. 
%
The above is always larger than 
$\varepsilon_{\rm min}$, where
\begin{equation}
\varepsilon_{\rm min}\sim\left(\frac{m_{3/2}}{m_P}\right)^{\frac{n}{n+1}},
\label{epsmin}
\end{equation}
where we also used Eq.~(\ref{epsbound}) with \mbox{$H_*\sim m_{3/2}$}.

Let us now enforce the constraint in
Eq.~(\ref{H*bound}), which, for \mbox{$H_*\sim m_{3/2}$}, reads
\begin{equation}
\varepsilon<\frac{\sqrt{\Omega_{\rm dec}}}{\pi\zeta}
\left(\frac{m_P}{T_{\sc bbn}}\right)^{1/2}
\left(\frac{m_{3/2}}{m_P}\right)^{5/4}\sim\;
10^{-4}\sqrt{\Omega_{\rm dec}}\;.
\label{epsbound0}
\end{equation}
From Eqs.~(\ref{epszeta}) and (\ref{epsbound0}) it is easy to find that
the above bound can be satisfied only if $n$ is large enough:
\begin{equation}
n>
\frac{8+\log(\sqrt{\Omega_{\rm dec}}/\theta)}{7
-\log(\sqrt{\Omega_{\rm dec}}/\theta)}.
\label{nbound}
\end{equation}
According to Eq.~(\ref{wmap}), we see that, at the best of cases,
(when \mbox{$\Omega_{\rm dec}\sim 10^{-2}$} and \mbox{$\theta\sim 1$}) we have 
\mbox{$n\geq 1$}. Hence, we see that {\em the radial field must correspond to 
a flaton field, stabilised by non-renormalisable terms}.
An upper bound on $n$ can be obtained by requiring that the curvaton
decays before BBN. 

The decay of the curvaton depends on its coupling to other particles. The
lowest decay rate corresponds to gravitational decay with 
\mbox{$\Gamma_\sigma\sim m_\sigma^3/m_P^2$}. However, if $\phi$ is part of a
supersymmetric theory we may expect a much larger value for $\Gamma_\sigma$.
An interesting possibility is realised by introducing the following coupling 
between $\phi$ and the Higgses:
\begin{equation}
\Delta W=\lambda_h\frac{\phi^{n+1}}{m_P^n}h^2
\label{DW}
\end{equation}
In this case, as is evident from Eq.~(\ref{fPQ}), our curvaton model also 
solves the $\mu$-problem for \mbox{$\lambda_h/\lambda\sim 1$}.

Now, the interaction of $\sigma$ with ordinary particles is governed by the 
effective $\mu$--term in Eq.~(\ref{DW}), which results into the following decay
rate of $\sigma$ into two Higgs particles:
\begin{equation}
\Gamma_\sigma\simeq\frac{(n+1)^2}{4\pi}\frac{m_\sigma^3}{v_0^2}.
\label{Gs}
\end{equation}
Demanding that \mbox{$\Gamma_\sigma\geq H_{\sc bbn}$} results in the bound
\begin{equation}
\Gamma_\sigma\sim
10^{-\frac{30n}{n+1}}
\left(\frac{m_\sigma}{\rm TeV}\right)^3{\rm TeV}
\geq H_{\sc bbn}\sim 10^{-27}{\rm TeV}
\qquad\Rightarrow\qquad 
m_\sigma\geq 10^{\frac{n-9}{n+1}}{\rm TeV}\;,
\label{nboundup}
\end{equation}
where we used Eq.~(\ref{fPQ}). Since we consider 
\mbox{$m_\sigma\sim m_{3/2}\lsim$ TeV} we see that there is a mild upper bound 
on $n$
which, roughly, demands \mbox{$n\lsim 9$}.

To proceed further, we have to consider separately the cases when the 
curvaton decays before or after it dominates the Universe.
%
Suppose, at first, that the curvaton decays before domination 
($\Omega_{\rm dec}\ll 1$). In this case, 
during the radiation epoch and after the onset of the 
oscillations, for the curvaton density fraction we have 
\mbox{$\rho_\sigma/\rho\propto a(t)\propto H^{-1/2}$}.
Hence, 
we find
\begin{equation}
\Omega_{\rm dec}\sim
\left(\frac{\min\{m_\sigma, \Gamma_{\rm inf}\}}{\Gamma_\sigma}\right)^{1/2}
\left(\frac{\sigma_{\rm osc}}{m_P}\right)^2,
\label{bla}
\end{equation}
where we have used Eq.~(\ref{sosc}) and 
\begin{equation}
\left.\frac{\rho_\sigma}{\rho}\right|_{\rm osc}\sim
\left(\frac{\sigma_{\rm osc}}{m_P}\right)^2,
\label{rhofracosc}
\end{equation}
with
\mbox{$(\rho_\sigma)_{\rm osc}\simeq\frac{1}{2}m_\sigma^2\sigma_{\rm osc}^2$} 
and \mbox{$\rho_{\rm osc}\sim m_\sigma^2m_P^2$}.
Using Eq.~(\ref{Gs}) into Eq.~(\ref{bla}) and also Eqs.~(\ref{fPQ}) and 
(\ref{sosc}) 
we obtain
\begin{equation}
\varepsilon\sim\frac{\sqrt{g\Omega_{\rm dec}}}{\pi\zeta}
\left(\frac{m_{3/2}}{m_P}\right)^{\frac{1}{2}(\frac{n+2}{n+1})},
\label{epszeta1}
\end{equation}
where we have also used that
\mbox{$\Gamma_{\rm inf}<H_*\sim m_{3/2}\sim m_\sigma$} and
\begin{equation}
\Gamma_{\rm inf}\sim g^2m_{3/2}\;,
\label{Ginf}
\end{equation}
with the mass of the inflaton field $s$ taken to be 
\mbox{$m_s\lsim H_*\sim m_{3/2}$} and $g$ being the coupling of the 
inflaton to its decay products. In principle, $g$ can be as low as 
$m_s/m_P$ if the inflaton decays gravitationally. However, since 
reheating has to occur before BBN, $g$ has to lie in the range:
\begin{equation}
10^{-14}\sim 10\;\frac{m_{3/2}}{m_P}<g<1.
\label{grange}
\end{equation}
Combining Eqs.~(\ref{epszeta}) and (\ref{epszeta1}) we find the relation
\begin{equation}
\frac{g}{\Omega_{\rm dec}}\sim\frac{1}{\theta^2}
\left(\frac{m_{3/2}}{m_P}\right)^{\frac{n-2}{n+1}}.
\label{gW1}
\end{equation}


Let us now consider the case when the curvaton decays after domination 
($\Omega_{\rm dec}\approx 1$). In this case, 
the curvaton dominates the energy density of the Universe when
\mbox{$H=H_{\rm dom}$}, where $H_{\rm dom}$ is given by 
\begin{equation}
H_{\rm dom}\sim\left(\frac{\sigma_{\rm osc}}{m_P}\right)^4
\min\{m_\sigma, \Gamma_{\rm inf}\}\;.
\label{Hdom}
\end{equation}
Now, using Eqs.~(\ref{soscv0}), (\ref{Gs}) and (\ref{Ginf}) it can be shown 
that the requirement \mbox{$\Gamma_\sigma<H_{\rm dom}$} results in the bound
\begin{equation}
g>\frac{1}{\theta^2}\left(\frac{m_{3/2}}{m_P}\right)^{\frac{n-2}{n+1}}.
\label{gW2}
\end{equation}
The fact that the case of curvaton domination requires a larger value of $g$ 
[compare the above with Eq.~(\ref{gW1})] is to be expected because, this means 
that the inflaton 
decays earlier and, therefore, the density fraction $\rho_\sigma/\rho$ grows 
substantially, allowing the latter to dominate the Universe before its decay. 
The higher $g$ is the more dominant the curvaton will be.

Note also, that, when the curvaton decays after it dominates the Universe, 
the hot big bang begins after curvaton decay, which suggests the reheating 
temperature
\begin{equation}
T_{\sc reh}\sim\sqrt{\Gamma_\sigma m_P}\sim
m_{3/2}\left(\frac{m_{3/2}}{m_P}\right)^{\frac{1}{2}(\frac{n-1}{n+1})}.
\label{Treh1}
\end{equation}
It can be easily checked that the above is
higher that $T_{\sc bbn}$ when \mbox{$n\leq 9$}, in agreement with 
the bound from Eq.~(\ref{nboundup}).

From Eqs.~(\ref{gW1}) and (\ref{gW2}) we see that, in general,
\begin{equation}
g\geq\frac{\Omega_{\rm dec}}{\theta^2}
\left(\frac{m_{3/2}}{m_P}\right)^{\frac{n-2}{n+1}}.
\label{n}
\end{equation}
For \mbox{$\theta\sim 1$} and in view of Eqs.~(\ref{wmap})
and (\ref{grange}) the above bound suggests
\begin{equation}
n\geq 2\,,
\label{nrange}
\end{equation}
which is tighter than the bound in Eq.~(\ref{nbound}).

%
%

We now concentrate of the evolution of the radial field $|\phi|$, which
has to be such, as to achieve the required value for $\varepsilon$.
%
%
Let us assume, at first, that the radial field follows the growth of the 
temporal minimum given by Eq.~(\ref{vexact}). In this case we can calculate 
the amplification factor as
\begin{equation}
\varepsilon\equiv\frac{(|\phi|_{\rm min})_*}{v_0}=
\left[1-C_\phi\left(\frac{H_*}{m_\phi}\right)^2\right]^{\frac{1}{2(n+1)}}.
\label{eps1}
\end{equation}
Using Eqs.~(\ref{mcurv}) and (\ref{vexact}) one finds that, in this case,
the curvaton's mass is given by
\begin{equation}
\tilde{m}_\sigma^2\approx(C_A H+A)\sqrt{m_\phi^2-C_\phi H^2}
\end{equation}
The above and (\ref{eps1}) suggest that
\begin{equation}
\left(\frac{\tilde{m}_\sigma}{H_*}\right)^2
\simeq(A+C_AH_*)\frac{m_\phi}{H_*^2}\;\varepsilon^{n+1},
\end{equation}
which agrees with Eq.~(\ref{mH}), given that \mbox{$C_A\sim 1$} and
\mbox{$H_*\sim A\sim m_\phi\sim m_\sigma\sim m_{3/2}$}.
From Eq.~(\ref{vexact}) it is easy to show that the rate of growth of the
order parameter is
\begin{equation}
\frac{\dot{v}}{v}=\frac{|\dot{\phi}|_{\rm min}}{|\phi|_{\rm min}}=
\frac{\epsilon}{n+1}\left(\frac{m_\phi^2}{C_\phi H^2}-1\right)^{-1}H\,,
\label{vdotv}
\end{equation}
where $\epsilon$ is defined in Eq.~(\ref{vareps}).
From Eqs.~(\ref{eps1}) and (\ref{vdotv}) we obtain
\begin{equation}
(\dot{v}/v)_*\sim\epsilon_*\,\varepsilon^{-2(n+1)} H_*\,.
\label{vdv}
\end{equation}
Comparing this with Eq.~(\ref{nsbound}), we
find that, for the scale invariance of the spectrum to be preserved, we require
\begin{equation}
\epsilon_*\ll \varepsilon^{2(n+1)},
\label{ee0}
\end{equation}
where \mbox{$\epsilon_*=\epsilon(s_*)$}.

Now, if the growth of $|\phi|_{\rm min}$ is so rapid that the radial field 
cannot follow it, then  we expect $|\phi|$ to roll, instead, down the potential
hill. In this case the order parameter is determined by the rolling $|\phi|$.

When the cosmological scales exit the horizon the radial field has to be 
slowly rolling because we need the order parameter to vary slowly enough, not
to destabilise the approximate scale invariance of the perturbation spectrum 
[cf. Eq.~(\ref{nsbound})].
Therefore, the Klein-Gordon equation for $|\phi|$ is: 
\begin{equation}
3H_*|\dot{\phi}|-\bar m_\phi^2|\phi|\simeq 0\,,
\label{KG}
\end{equation}
where
\begin{equation}
\bar m_\phi^2\equiv m_\phi^2-C_\phi H^2.
\label{mS}
\end{equation}
Using the above, the rate of growth of the order parameter, in this case, 
can be easily found to be
\begin{equation}
\frac{\dot v}{v}=\frac{|\dot\phi|}{|\phi|}
=\frac{1}{3}C_\phi\left(\frac{m_\phi^2}{C_\phi H^2}-1\right)H\,.
\label{fdotf}
\end{equation}

The variation of the order parameter is expected to follow the less rapidly
changing rate of growth. Hence, by comparing the two rates in 
Eqs.~(\ref{vdotv}) and (\ref{fdotf}), we see that the order parameter follows 
the variation of $|\phi|_{\rm min}$ only if
\begin{equation}
\varepsilon^{4(n+1)}>
\epsilon_*\;,
\label{ee}
\end{equation}
where we used again Eq.~(\ref{eps1}) and also \mbox{$C_\phi\sim 1$}.
It is evident that, if the above constraint is satisfied then
so is the requirement in Eq.~(\ref{ee0}).
Note, however, that if the above constraint is
violated then the order parameter $v$ is determined 
by the rolling $|\phi|$ and not by the varying 
$|\phi|_{\rm min}$, in which case the requirement 
in Eq.~(\ref{ee0}) is not valid, while also
the amplification factor is not the one shown in Eq.~(\ref{eps1}).%
\footnote{If $v$ follows the growth of $|\phi|$
instead of $|\phi|_{\rm min}$ then $\varepsilon$
is expected to be smaller that the one in 
Eq.~(\ref{eps1}) because, at any given time,
\mbox{$v(t)<|\phi|_{\rm min}(t)$}.}

In this latter case, we find the amplification factor 
as follows. Using Eq.~(\ref{VN}) we can write
$|\phi|$ as a function of the number $N$ of the 
remaining e-foldings of inflation. Starting from
Eq.~(\ref{KG}) and after a little algebra we obtain
\begin{equation}
\frac{3}{C_\phi}\frac{d\ln|\phi|}{dN}=
\frac{e^{-2F_sN_{\rm x}}-e^{-2F_sN}}{1-e^{-2F_sN}},
\label{dSN}
\end{equation}
where $N_{\rm x}$ corresponds to the phase transition
which changes the sign of $\bar m_\phi^2$. By definition
\begin{equation}
m_\phi^2\equiv C_\phi H^2_{\rm x}\simeq
C_\phi H_{\rm inf}^2(1-e^{-2F_sN_{\rm x}}).
\label{Hx}
\end{equation}
where \mbox{$H_{\rm x}\equiv H(N_{\rm x})$}.
Integrating Eq.~(\ref{dSN}) we get
\begin{equation}
\frac{6}{C_\phi}\ln\left(\frac{|\phi|_*}{|\phi|_{\rm x}}\right)=
(1-e^{-2F_sN_{\rm x}})F_s^{-1}
\ln\left|\frac{e^{2F_sN_{\rm x}}-1}{e^{2F_sN_*}-1}\right|
-2(N_{\rm x}-N_*),
\label{SN}
\end{equation}
where \mbox{$|\phi|_{\rm x}\equiv|\phi|(N_{\rm x})$}.
%

The displacement of the field from the origin at the phase 
transition is determined by its quantum fluctuations. This means that
\begin{equation}
|\phi|_{\rm x}\simeq H_{\rm x}/2\pi\,,
\label{SN0}
\end{equation}
We also have
\begin{equation}
\varepsilon=\frac{|\phi|_*}{v_0}=\frac{|\phi|_*}{H_*}
\frac{H_*}{v_0}\Rightarrow
|\phi|_*\simeq
\frac{\varepsilon}{\varepsilon_{\rm min}}
\;\frac{H_*}{2\pi}\,,
\label{See}
\end{equation}
where we used Eq.~(\ref{epsbound}).


In view of Eq.~(\ref{fdotf}),
the requirement in Eq.~(\ref{nsbound}) becomes
\begin{equation}
\frac{3}{C_\phi}e^{-2F_sN_*}\left(\frac{1-e^{-2F_s(N_{\rm x}-N_*)}}%
{1-e^{-2F_sN_*}}\right)\ll 1\,.
\label{nnn}
\end{equation}
where we took into account Eq.~(\ref{Hx}). 


Finally, another issue to be addressed concerns the
requirement that the radial field {\em does} slow roll
at the time when the cosmological scales exit
the horizon.
In order for this to occur, its quantum fluctuations should not dominate its 
motion, i.e. $|\phi|$ has to be outside the quantum
diffusion zone. The condition for this to occur is 
\mbox{$H_*/2\pi<(\dot\phi/H)_*$} or equivalently
\begin{equation}
\left.\frac{\partial V}{\partial|\phi|}\right|_*
\simeq 2\bar m_\phi^2(H_*)|\phi|_*>H_*^3\,.
\end{equation}
Using Eqs.~(\ref{mS}) and (\ref{Hx}) and working
as before, the above constraint is recast as
\begin{equation}
\ln\left(\frac{|\phi|_*}{|\phi|_{\rm x}}\right)>2F_sN_*-\ln(C_\phi/\pi)+
\ln\left(\frac{1-e^{-2F_sN_*}}{1-e^{-2F_s(N_0-N_*)}}\right)
+\frac{1}{2}\ln\left(\frac{1-e^{-2F_sN_*}}{1-e^{-2F_sN_{\rm x}}}\right).
\label{SNbound}
\end{equation}

%
%

To illustrate the above we present an example, taking 
\begin{equation}
n=2\quad{\rm and}\quad\lambda,\theta\sim 1\,. 
\end{equation}
The bound in Eq.~(\ref{nboundup}) suggests that this is 
acceptable provided \mbox{$m_\sigma\gsim$ 5~GeV}. 
Using Eq.~(\ref{epszeta}) we obtain the value
of the amplification factor, necessary for the
model to work:
\begin{equation}
\varepsilon\sim 10^{-6}\Omega_{\rm dec}\,.
\label{e-value}
\end{equation}

If the curvaton decays after domination 
then Eq.~(\ref{gW2}) demands \mbox{$g>1$}, which is not compatible with the
range in Eq.~(\ref{grange}). Therefore, we have to assume that the curvaton 
decays before domination, in which case \mbox{$\Omega_{\rm dec}\lsim 1$}, with 
the bound saturated when the curvaton decays approximately when it is about to 
dominate the Universe. In this case, Eq.~(\ref{gW1}) suggests
\begin{equation}
g\sim\Omega_{\rm dec}\lsim 1\,.
\label{g-value} 
\end{equation}
Such a large coupling can be understood only if the VEV of the inflaton modulus
is an enhanced symmetry point. As a result of the above, the reheating 
temperature after the end of inflation is found to be
\begin{equation}
T_{\rm reh}\sim g\sqrt{m_{3/2}m_P}\sim g\times 10^{10.5}\;{\rm GeV}.
\end{equation}
From the above we see that, in order not to challenge the gravitino constraint,
we have to choose the lowest possible value of $g$, which, according to 
Eqs.~(\ref{wmap}) and (\ref{g-value}) corresponds to
\begin{equation}
\Omega_{\rm dec}\sim 10^{-2}.
\end{equation} 
Hence, from Eqs.~(\ref{e-value}) and (\ref{g-value}) we obtain the values
\begin{equation}
\varepsilon\sim 10^{-8}\quad{\rm and}\quad g\sim 10^{-2}.
\label{eg}
\end{equation}
From Eq.~(\ref{s0}) and (\ref{H0}) and also, 
using Eq.~(\ref{vareps}), it is easy to 
see that
\begin{equation}
\epsilon_*>10^{-30},
\label{vareps1}
\end{equation}
where we considered that \mbox{$s_*>s_{\rm in}$}.
Hence, from Eqs.~(\ref{eg}) and (\ref{vareps1})
it is straightforward to see that the constraint in Eq.~(\ref{ee}) is badly
violated, which means that the order parameter $v$ follows the slow roll of the
$|\phi|$ field and not the variation of the minimum of the potential 
$|\phi|_{\rm min}$. Consequently, the amplification factor 
$\varepsilon$ is {\em not} given by the expression in Eq.~(\ref{eps1}) in this 
case. Instead, we can estimate the amplification
factor with the use of Eq.~(\ref{See}). 

Using Eq.~(\ref{epsmin}) with \mbox{$n=2$} we
find 
\begin{equation}
\varepsilon_{\rm min}\sim 10^{-10}.
\end{equation}
Therefore, Eqs.~(\ref{epsmin}), (\ref{See}) and (\ref{eg}) suggest
\begin{equation}
|\phi|_*\sim 10^2\frac{H_*}{2\pi}\,.
\label{cond}
\end{equation}
The above can, in principle, be used in Eqs.~(\ref{SN}) and (\ref{SNbound})
to constrain the parameters of the underlying model. 

A useful quantity to calculate in order to 
evaluate Eqs.~(\ref{SN}) and (\ref{SNbound}) is 
the number of e-foldings, which corresponds to 
the cosmological scales $N_*$. The cosmological
scales range from a few times the size of the 
horizon today $\sim H_0^{-1}$ down to 
scales $\sim 10^{-6}H_0^{-1}$ 
corresponding to
masses of order $10^6M_\odot$ \cite{book}. Typically this
spans about 13 e-foldings of inflation. For the
estimate of $N_*$ we will chose a scale roughly 
in the middle of this range; the scale that 
re-enters the horizon at the time when structure
formation begins, i.e. at the time $t_{\rm eq}$ 
of matter--radiation equality. Then, in the case when the curvaton decays 
before domination it is straightforward to obtain
\begin{equation}
\exp(N_*)\sim H_*^{1/3}\Gamma_{\rm inf}^{1/6}\sqrt{t_{\rm eq}}
\sim g^{1/6}\sqrt{m_{3/2}t_{\rm eq}}
\end{equation}
where we have used Eq.~(\ref{Ginf}) and that 
\mbox{$H_*\sim m_\sigma\sim m_{3/2}$}.
Using Eq.~(\ref{eg}), we obtain
\begin{equation}
N_*\simeq 43\,.
\label{N*}
\end{equation}
The number of e-folds that corresponds to decoupling 
(when the CMBR is emitted) is
roughly \mbox{$N_*+1.5$}, while the one which corresponds to 
the present horizon is \mbox{$\sim N_*+9$}.

In the attempt to obtain the allowed parameter space for our model
it soon becomes clear that, while the requirement in Eq.~(\ref{SNbound})
is relatively easy to satisfy, the major difficulty is reconciling 
Eq.~(\ref{SN}) with the bound in Eq.~(\ref{nnn}) coming from the spectral index
requirements. This is especially true in view of the recent WMAP results 
\cite{wmap1}, which correspond to spectral index \mbox{$n_s=0.96\pm 0.02$}, 
i.e. \mbox{$n_s\geq 0.92$} at 95\% c.l. This means that the left-hand-side of
Eq.~(\ref{nnn}) should not exceed 0.04.\footnote{Note that, in our model, all 
other contributions to the deviation of the spectral index from unity 
\cite{trotta} are negligible.} By careful investigation of 
Eqs.~(\ref{SN}) and (\ref{nnn}) it is found that the above difficulty is 
more alleviated the larger the value of $N_{\rm x}$ is, i.e. the earlier the
phase transition occurs. In fact, a solution is only possible if 
\begin{equation}
2F_sN_{\rm x}\gg 1\,.
\label{approx}
\end{equation}

In view of the above Eqs.~(\ref{SN}) and (\ref{nnn}) can be respectively
approximated as
\begin{eqnarray}
\ln\left(\frac{|\phi|_*}{|\phi|_{\rm x}}\right)
& \simeq & -\frac{C_\phi}{6F_s}\ln(1-e^{-2F_sN_*})
\label{fN0}\\
C_\phi & \leq & 0.12\,(e^{2F_sN_*}-1)\,.
\label{cbound}
\end{eqnarray}
Now, using Eq.~(\ref{SN0}) we can write:
\begin{equation}
\frac{|\phi|_*}{|\phi|_{\rm x}}\simeq\frac{2\pi|\phi|_*}{H_*}
\left(\frac{H_*}{H_{\rm x}}\right)\;\Rightarrow\;
\ln\left(\frac{|\phi|_*}{|\phi|_{\rm x}}\right)\simeq
\ln\left(\frac{2\pi|\phi|_*}{H_*}\right)
+\frac{1}{2}\ln\left(\frac{1-e^{-2F_sN_*}}{1-e^{-2F_sN_{\rm x}}}\right),
\label{rid}
\end{equation}
where we have considered also Eq.~(\ref{VN}), using that 
\mbox{$H^2(N)\simeq V(N)/3m_P$}.
In view of the above and according to the approximation in Eq.~(\ref{approx})
we can recast Eq.~(\ref{fN0}) as
\begin{equation}
\ln\left(\frac{2\pi|\phi|_*}{H_*}\right)
\simeq -\left(\frac{1}{2}+\frac{C_\phi}{6F_s}\right)\ln(1-e^{-2F_sN_*}).
\label{fN}
\end{equation}
Under the same approximation Eq.~(\ref{SNbound}) becomes
\begin{equation}
\ln\left(\frac{2\pi|\phi|_*}{H_*}\right)>2F_sN_*-\ln(C_\phi/\pi)+
\ln(1-e^{-2F_sN_*})\,,
\label{fNbound}
\end{equation}
where we have also used Eq.~(\ref{rid}).

Solving Eq.~(\ref{fN}) in terms of $C_\phi$ and using Eq.~(\ref{cbound}) 
we obtain
\begin{equation}
-\frac{2F_sN_*}{86}\left[1+\frac{4\ln 10}{\ln(1-e^{-2F_sN_*})}\right]-
0.04(e^{2F_sN_*}-1)\leq 0
\end{equation}
where we have also employed Eqs.~(\ref{cond}) and (\ref{N*}). Solving 
numerically we obtain the bound
\begin{equation}
F_s\lsim\frac{1}{560}\simeq 1.8\times 10^{-3}
\label{Fbound}
\end{equation}
This bound, in view of Eq.~(\ref{Fs}) results in 
\begin{equation}
m_s\leq 0.073\,H_{\rm inf}\;.
\label{msbound}
\end{equation}
which is somewhat tight and implies that inflation is not really of the 
fast-roll type, but the inflaton is light enough to roll slowly down its
potential hill. From Eqs.~(\ref{Ntot}) and (\ref{Fbound}) one obtains
\begin{equation}
N_{\rm tot}\geq 1.9\times 10^4.
\end{equation}
Thus, if the phase transition, which releases $|\phi|$ from the origin,
occurs not much later than the onset of inflation, then the approximation in 
Eq.~(\ref{approx}) can be well justified.
Similarly, using Eqs.~(\ref{N*}) and (\ref{Fbound}), Eq.~(\ref{cbound}) 
gives the bound
\begin{equation}
C_\phi\leq 0.020\,.
\label{Cbound}
\end{equation}
In view of Eqs.~(\ref{Hx}) and (\ref{approx}) the above bound suggests
\begin{equation}
m_\phi\leq 0.14\,H_{\rm inf}\,.
\label{mfbound}
\end{equation}
The values of $m_s$ and $m_\phi$ can approach $H_{\rm inf}$
if one decreases $N_{\rm x}$ but then the constraint 
in Eq.~(\ref{nnn}) becomes seriously challenged. 
It can be easily checked that, with the above values the requirement
in Eq.~(\ref{fNbound}) is satisfied as well.


The above results 
suggest that, for the $n=2$ case and 
when \mbox{$\lambda,\theta\sim 1$}, the model can work for masses of the order
\begin{equation}
m_\phi\lsim 0.1\,
m_{3/2}
\quad{\rm and}\quad
m_s\lsim 0.01\,
m_{3/2}
\,,
\end{equation}
where \mbox{$
m_{3/2}\sim 1$~TeV}.
Such values imply only a mild tuning on the masses; predominantly on the
mass of the inflaton modulus. This is necessary because the variation of $H$ 
should be kept small, since only then can the tachyonic effective mass of the 
radial field $\bar m_\phi$ remain small enough for $|\phi|$ to be slow-rolling 
and the constraint in Eq.~(\ref{nsbound}) to be satisfied. Note that, a tuning
of the inflaton mass is quite plausible, since the latter is a string axion.

One may wonder why, since both the inflaton 
field $s$ and the radial field $|\phi|$ turn-out to be 
light when the cosmological scales exit the 
horizon during inflation, we cannot use those 
fields to generate the observed curvature perturbations. 
The reason is that, in contrast to the PNGB 
curvaton, the perturbations of those fields are
not amplified. Hence their contribution to the
overall curvature perturbation is insignificant.
Indeed, for the inflaton we have 
\mbox{$\zeta_s\sim(m_s/s_*)\sim 
10^{-17}$}, which is much 
smaller than the observed value \mbox{$\zeta\simeq 2\times 10^{-5}$}.
Similarly, for $|\phi|$ it is easy to show that 
\mbox{$\zeta_\phi\sim\varepsilon\zeta_\sigma\sim 10^{-13}$},
where we used that
\mbox{$\zeta_\sigma\approx\zeta$}.


In conclusion we have seen that our mechanism can work with natural values
of the parameters with only a mild tuning on the inflaton mass. Another
important requirement is that the phase transition, which releases the radial 
field from the origin, occurs much earlier than the time when the cosmological 
scales exit the horizon, in order not to destabilise the flatness of the 
curvature perturbation spectrum.

Our PNGB curvaton is such that can be easily accommodated in simple extensions 
of the standard model. Indeed, in Ref.~\cite{ours} we present in detail such a 
realisation, using as curvaton an angular degree of freedom orthogonal to the 
QCD axion in a class of supersymmetric constructions of the Peccei-Quinn 
symmetry. Presumably, other PNGB curvatons, such as the ones in 
Ref.~\cite{pngb}, can also be utilised.

We should note here that, although the modular inflation model, which we 
considered, is highly motivated, it is by no means the only possibility. 
Other inflationary models with Hubble-scale of order 1~TeV 
may also be applied \cite{minos}. Needless to say that designing 
inflationary models at such energy scale can allow direct contact with 
particle physics.

\bigskip

I am grateful to G.~Lazarides 
for stimulating discussions.

\end{document}